\documentstyle[11pt,aaspp4]{article}

\begin{document}

\title{Eclipsing Binaries in the OGLE Variable Star Catalog.~IV.\\
The Pre-Contact, Equal-Mass Systems}

\author{\sc Carla Maceroni\\
\rm Electronic-mail: {\it maceroni@coma.mporzio.astro.it\/}}
\affil{Rome Observatory,
via Frascati 33, I-00040 Monteporzio C., Italy
}

\and

\author{\sc Slavek M. Rucinski\\
\rm Electronic-mail: {\it rucinski@astro.utoronto.ca\/}}
\affil{David Dunlap Observatory, University of Toronto \\
P.O.Box 360, Richmond Hill, Ontario, Canada L4C~4Y6}

\bigskip
\centerline{\today}

\bigskip
\begin{abstract}
We used the database of eclipsing binaries detected
by the OGLE microlensing project in the pencil-beam search volume
toward Baade's Window to define a
sample of 74 detached, equal-mass, main-sequence binary stars
with short orbital periods in the range $0.19 < P < 8$ days.
The logarithmic slope of the period distribution,
$\log N \propto (-0.8 \pm 0.2)\, \log P$, was used to infer the
angular-momentum-loss (AML) efficiency for the late,
rapidly-rotating members of close binaries.
It is very likely that the main cause of the negative slope is a discovery
selection bias that progressively increases with the
orbital period length. Assuming  a power-law dependence
for the correction for the bias $\propto -C \log P$ (with $C \ge 0$)
the AML braking-efficiency exponent $\alpha$ in $dH/dt = P^{-\alpha}$
can take any value $\alpha = -1.1\, (\pm 0.2) + C$.
Very simple considerations of discovery biases suggest
$C \simeq 4/3$, which
would give an AML braking law very close to the ``saturated'' one,
with no dependence on the period. However, except for plausibility
arguments, we have no firm data to support this estimate of $C$,
so that $\alpha$ remains poorly constrained. The results
signal the utmost importance of  the detection bias evaluation
for variable star databases used in analyses similar to the
one presented in this study.
\end{abstract}

\section{INTRODUCTION}
\label{intro}
This paper is a continuation of the analysis of the eclipsing binaries
detected by the OGLE microlensing project in the nine central
fields of Baade's Window (BWC to BW8) toward the Galactic Bulge
(Udalski et al. \markcite{uda94}1994, \markcite{uda95a}1995a,
\markcite{uda95b}1995b).
The three previous papers of this series addressed the properties of
contact binaries which are the most common type in the sample
of 933 eclipsing systems in Baade's Window (BW). The first paper
(Rucinski \markcite{ruc97a}1997a = R97a) showed that -- due to their
high frequency of incidence -- the contact systems
of the W~UMa-type can be useful
distance indicators along the line of sight all the way to the
Galactic Bulge and that they belong to the old galactic disk
population. The second paper (Rucinski \markcite{ruc97b}1997b
= R97b) discussed the light curves of those systems.
The light-curve amplitude distribution strongly
suggested a mass-ratio distribution steeply climbing toward low
mass-ratios (i.e.\ unequal masses). The systems with unequal temperatures
of components, which are seen in the contact-binary sample
as a small admixture at the level of 2 percent,
in their majority are not poor-thermal-contact
systems but rather semi-detached binaries with
matter flowing from the hotter, more massive
component and forming an accretion hot spot on the cooler companion.
The third paper (Rucinski \markcite{ruc98a}1998a = R98a)
dealt with contact systems with orbital periods longer than one day.
The W~UMa-type sequence was shown to continue up to the orbital
periods of 1.3 -- 1.5 day, and then sharply terminate in this period
range.
The results of the three previous papers, R97a, R97b, R98a, were
re-discussed in a more general comparison of the contact binaries of
the Galactic Disk in the BW sample with those in old open clusters
(Rucinski \markcite{ruc98b}1998b = R98b). It was found that the
luminosity function for the contact binaries is
very similar in shape to that for the solar neighborhood
main-sequence (MS) stars, implying a flat apparent
frequency-of-occurrence distribution. In the accessible
interval $2.5 < M_V < 7.5$, the apparent frequency of contact binaries
relative to MS stars was found to be equal about 1/130 -- 1/100.
The resulting spatial (inclination-corrected) frequency of some 1/80
(with a combined uncertainty of about $\pm 50$ percent)
implies a well-defined and high peak in the orbital period distribution,
well above the period distribution for MS binaries by
Duquennoy \& Mayor \markcite{duq91} (1991), extrapolated
to periods shorter than one day. This peak most probably
results from piling-up of short-period binaries as they lose
angular momentum and form relatively long-lived contact systems.

This paper is an attempt to extend  the analysis of the eclipsing
systems discovered by OGLE into the pre-contact domain. In utilizing
the observed period distributions, it is
logically related  to the studies on
the  orbital period evolution of tidally-locked late-type
main-sequence binaries by angular
momentum loss (AML) that  were published in a series of papers
by Maceroni and Van't Veer (1989, 1991 =MV91 \markcite{MV91}; see also
see Maceroni (1999) \markcite{CM99}  for an update and further references).
In particular, a functional form of the AML rate was
derived in MV91 and Maceroni (1992) \markcite{CM92}  by fitting the observed
period distribution of field binaries.
The results nicely confirmed the need of a braking
mechanism which is weakly rotation-dependent
(or, perhaps, totally independent of the rotation rate) for fast rotators,
but the conclusions from these analyses suffered from the unavoidable
inhomogeneity of the all-sky sample.
A subsequent study by St\c epie\'n (1995) \markcite{Ste95}
arrived at basically the same results through a very different route,
via specific assumptions on the efficiency of the
magnetic-field generation and an analysis of the
homogeneous (but very small) sample of the Hyades binaries.
Attempts to relate these theoretical predictions to the statistics
of short-period binaries  have so far been encountering severe
limitations related to the smallness of the samples.
We note that early indications that the orbital
period evolution at very short periods
is slower than initially expected, based on
bright field-star binary statistics, were presented by Rucinski
\markcite{ruc83} (1983), but 
suffered as well from low-number statistical uncertainty.
The currently on-going microlensing surveys offer, for the first time,
rich and homogeneous samples of binaries to overcome those
limitations.

This paper uses the sample of eclipsing binaries observed
by the OGLE project in the direction
of Baade's Window sample. The sample is of moderate size
(933 systems) by the rapidly-evolving standards of
the microlensing projects,
yet it remains the only widely-available sample of that type.
As we explain in Section~\ref{sample},
the discussion is limited -- by necessity -- to eclipsing
binaries with almost equally-massive components. This again
limits the size of the available sample. Although
the results are tentative, we decided to present them for completeness
and as a guidance for the future, larger surveys.

Section~\ref{aml} very briefly summarizes the expected trends for
the pre-contact domain while Section~\ref{sample} contains the definition
of the sample and its properties. Section~\ref{dist} presents analysis
of the observed period distribution. The last Section~\ref{disc}
summarizes the main results of the paper.

\section{ANGULAR-MOMENTUM-LOSS EVOLUTION OF CLOSE BINARY SYSTEMS}
\label{aml}

The evolution of a close binary system crucially depends on its orbital
period. When the period is short enough for an effective
tidal synchronization, the angular momentum lost by the individual
components through the action of magnetized wind is extracted from
the orbit. The orbital separation shrinks, the components rotate
progressively
faster, possibly draining even more angular momentum from the orbit.
Eventually, a contact system forms as a penultimate stage before
merging of components and formation of a single star.
This general description has been first suggested by
Van't Veer \markcite{vv79} (1979), later developed by
Vilhu \markcite{vil82} (1982), and then explored in more detail by
Maceroni and Van't Veer  \markcite{mv91} (1991)
and by St\c epie\'n \markcite{ste95} (1995).

The implications of the angular momentum loss (AML)
evolution are most obviously noticeable in the orbital
period distribution. The crucial quantities here are
the moment of inertia of
the layers effectively braked during this process and
the {\it rate\/} of the AML. The dependence of the AML rate
on the rotation period, $P_{rot}$, or on the stellar angular velocity of
rotation, $\omega = 2\pi/P_{rot}$, is frequently called the braking
``law'',
and is written as variants of $\dot{\omega}=\dot{\omega}(\omega)$ or
$\dot{P}_{rot} = \dot{P}_{rot}(P_{rot})$.
Frequently, rigid body rotation of the whole star is assumed as
this simplifies derivation of the moment of inertia of the star.
The detailed models of  MV91 show that the tidal synchronization
mechanisms operate on such a short time scale compared to the AML one,
that the hypothesis of perfect synchronization is fully
justified after a few million years of evolution, so that one can
write  $P_{rot}=P_{orb} \equiv P$,  dropping the suffixes.

The braking law is usually expressed in a parametric form of the
type: $\dot{\omega}= const \cdot  \omega^\alpha$, with $\alpha=3$
for the well know Skumanich relation (Skumanich \markcite{Sk72} 1972)
which is known to be valid for slowly-rotating, single, solar-type
stars.
In a perfect synchronization regime the angular momentum loss by
magnetic braking $\dot{H} \propto P^{-\alpha}$ is
equal to the decrease of the orbital angular momentum
$ \dot{H}_{orb} \propto P^{-2/3} \dot{P}$, so that the rate
of change of the orbital period is $\dot{P}\propto P^{2/3-\alpha}$.
The period distribution is then expected to be dependent on time,
with the population of the period-distribution bins directly
related to the shape of the initial distribution and to the time scale
of
period change.

The initial period  distribution  is  poorly
known in the short-period range of interest here. Current assumptions
are extrapolations towards short periods
either of the  $\log P$ flat distribution
of Abt and and Levy \markcite{AL76} (1976) (as for instance in MV91),
or  of the more recent one by
Duquennoy and Mayor \markcite{duq91} (1991 = DM91), that
is derived for a relatively unbiased sample.   Sometimes a
 short-period cut-off
has been introduced, as in St\c epie\'n \markcite{Ste95} (1995).
The DM91 distribution  has the
shape, in the logarithm of the period, of a wide Gaussian with a
maximum at $\log P = 4.8$ and $\sigma \log P = 2.3$
(with the period $P$ expressed in days); it can be approximated, within
 the range $ 0 < \log P < 1$ days, by $N(\log P) \propto P^{0.35}$.

The simplest case of evolution of the period  distribution is
that for a sample of systems all formed
at the same time $t_0$. If the initial   distribution is $f_0( \log P)$,
the evolved one  at a later time $t_*$ (e.g.\ the present time of
observation), $f_*(\log P)$, results from the requirement
of the constancy of the total number of the systems.
The implied  transformations of  both $f_0(\log P)$, and the bin
size ${\rm d} \log  P(t_0)$ are related through:
\begin {equation}
 f_*(\log  P )  = f_0(\log P) \left |  \frac{ {\rm d \log }
P(t_0) }{{\rm d}\log  P(t_*) }   \right |  =
 f_0(\log P) \frac{ \tau _*} { \tau _0} \label{1}
\end{equation}
which incorporates the period evolution from the time $t_0$ to $t_*$:
$P_*=P_*(P_0)$. For brevity, we use $P_n$ to signify $P(t_n)$.
The quantity $\tau=|P/\dot{P}| $ is the timescale of the period
evolution.
Thus, from Eq.\ \ref{1} we see that
the present period distribution function $f_*(\log P)$ is
proportional to the  ratio of the present and the
initial time-scales (i.e. rapid evolution will locally
deplete the distribution). Obviously,
in the more realistic case of the time-independent
formation process over some
time interval,  the present period distribution would be
a result of an integration of the right side of Eq.~\ref{1} over the
whole time interval.

The systems of our sample  cannot be considered strictly coeval:
The present population  of each  bin
presumably consists of binaries of somewhat different age
reaching the relevant bin by means of the AML occurring
since their formation.
On the other hand,  the stars in Baade's Window in their majority
probably belong to a relatively old population,
possibly older than $\sim 5$ Gyr (Ng et al.\ \markcite{ng97} 1996,
Kiraga et al.\ \markcite{kir97} 1997),
so that products of recent formation events are  quite unlikely to be
seen there.
 In that hypothesis, and according to all the numerical
models of period evolution we mentioned  before,
the systems presently observed as pre-contact binaries
come from a initial period range where the time-scale of period change
was
very long and weakly time dependent for the first few Gyr
of evolution. (Such a case
corresponds to the period evolution functions being
nearly straight and parallel  to the time axis,
as shown for instance in Figures 4 and 7 of MV91.)
 As a consequence the result of the integration over the  formation time
 will   be  proportional to   $f_0 \tau_*/\tau_0$,
where $\tau_0$ will be a mean over the formation time.
 As long as we  study  only the shape of the period
distribution we can  just
use the simpler expression in Equation \ref{1}.

According to the power law expressing the rate of orbital period change,
  the period evolution function relating $P_0$ to $P_*$  can be written
as:
\begin{displaymath}
P_0=P_* \left[1+\left(\alpha+1/3 \right)
\frac{T}{\tau_*}\right]^{1/(\alpha+1/3)}
\end{displaymath}
with $T= t_*-t_0$ and with $\tau_*$ being the present
time-scale of period evolution,
The present period distribution obtained from the initial distribution
$f_0(\log P)\propto P_0^{\beta}$ will be:
\begin{equation}\label{ppd}
f_*(\log P) \propto  P^{\beta}\left[1+\left(\alpha+1/3 \right) \frac{T}
{\tau_*} \right]^{\frac{\beta-\alpha-1/3}{\alpha +1/3} }
\end{equation}

\placefigure{fig1}

Figure~\ref{1} shows two examples of
the  evolution of an initial distribution
assumed to be as a power law $f(\log P)\propto P^\beta$ with
$\beta=0.35$ (the local fit of DM91),
for two values of $\alpha$, 1.49 and 3;
the first value corresponds to a local power-law
fit of the of St\c epie\'n \markcite{Ste95} (1995) braking relation ($-0.5< \log P
<1.0$), the second
is the  Skumanich's value \markcite{Sk72} (1972). In both cases
identical
solar-mass components were assumed.

The initial period ranges were assumed to be different for
each panel; they were adjusted to correspond to intervals of
initial periods that -- after 8 Gyr --
could populate the distribution down to 0.3 days,
the approximate value of the period for contact systems
consisting of solar components.
In the same way only the parts of the intermediate age distributions
that finally spread up to that lower boundary are shown.
As can  be seen from Equation \ref{ppd} the evolving
distribution, in a log--log
plot, gradually changes its slope  from  the
initial $\beta$ to the asymptotic value of $\alpha+1/3$.
 The speed of the process depends on the value of
time-scale $\tau$ (and not just on its slope with $\log P$) , this
in its turn depends on the stellar parameters. Again, for illustrative
purposes in Figure~\ref{1}, we have  used values derived 
from the St\c epie\'n and the Skumanich predictions.
Writing $\tau=k P^{\alpha+1/3}$ the  multiplicative constant $k$ turns out to be
$k \simeq 5.2$ and $k \simeq 0.5$  , respectively for the St\c epie\'n and the Skumanich
models ($\tau$ expressed in Gyr  and $P$ in days). The much shorter value of $\tau$ for the extrapolation
to short periods of the
Skumanich law gives a very rapid transition of
the slope to the asymptotic value,
and relatively long initial periods.
However even with the  reduced  braking efficiency at short periods of
the St\c epie\'n law, the short period part of the distribution very
soon loses
memory  of the initial period distribution.
We expect therefore  a relatively old population of binaries, as
that of
Baade's Window, to contain in the short period range of Figure~\ref{1}
 information  on the time-scale dependence
of $P$ rather than any vestiges of the initial distribution.

If the braking law were indeed as steep as implied by the Skumanich
relation
of Figure~\ref{1} ($\alpha = 3$), the AML evolution would
progressively accelerate with a rapid shortening of $P_{rot} = P_{orb}$
as  the time scale would decrease as
$\tau \propto P^{10/3}$
and there would be practically no very short-period,
synchronized binaries. For the solar-type dwarfs of spectral types FGK,
the region around orbital periods of about 0.3 to 2 days would become
quickly ``evacuated''. However, there are many indications that
$\alpha$ is definitely smaller than the Skumanich value
for  fast rotators, and that the braking law may,
in fact, ``saturate'', converging to a constant. Then, the rate of
the evolution would still depend on the period, but with a much
shallower
slope ($N \propto \tau \propto P^{1/3}$ for the limiting value of
$\alpha=0$) and longer $\tau$.

The strongest dependence on the AML efficiency is in any case
expected to take place at the very short end of the period
distribution. Presumably some temporary increase in the
number of very short-period systems that  formed with longer periods
may occur
there,
but -- over time - the distribution will reflect the AML efficiency as
the distribution evolves into the steady state determined by the braking
time-scale . At the very short period end one expects
a prominent effect of truncation as the close
detached systems become converted into contact binaries.
The final piling in the contact binary domain
is very clearly seen in the currently most extensive statistical
data for the Galactic Disk contact binaries visible in
the Baade's Window direction (R98b).

\placefigure{fig2}

Figure~\ref{fig2} compares
these data with the extrapolation of the DM91 period
distribution to very short periods. Validity of such an
extrapolation is put in question
by a small but homogenous sample of dwarfs in the young cluster of
Hyades
(Griffin \markcite{grif85} 1985) where an increase in star numbers
toward short periods within the range $1 < P < 10$ days
is actually observed (see Figure~\ref{fig2}).
Although the statistics for the Hyades binaries is poor,
the trend is quite obvious. It is not clear
if this is a remnant of the formation process or an indication
that the AML evolution actually {\it slows down\/} at very
short periods due to the inversion in the braking efficiency
(negative $\alpha$). Such a possibility is not entirely excluded
as the rapid-rotation regime is very poorly understood in terms
of the AML efficiency. For example, presence of thinly-populated
``tails'' of extremely rapidly-rotating late-type dwarfs
in young clusters (Hartmann \& Noyes \markcite{hn87} 1987)
may indicate such an inverted AML efficiency.
Besides, very close-binary stars do not have
to behave exactly as single stars due to the influence of the
tidal effects on the magnetic field generation modes.

In summary: The value of the exponent $\alpha$ in the braking
efficiency law, $\dot{H} \propto P^{-\alpha}$, for the high
rotation-rates in close binary systems is currently a very poorly
known quantity so that
any observational results which are free of systematic effects,
would be of great value. An estimate on $\alpha$
was the goal of the present paper.

\section{THE SAMPLE OF PRE-CONTACT SYSTEMS}
\label{sample}

\subsection{Definition of the sample}

The extraction of the sample of pre-contact systems is not a trivial
matter, given the limited photometric information  for the periodic
variable stars in the OGLE
catalog. The availability  of only single color light curves is
particularly inconvenient because --
without color curves -- we could not eliminate the semi-detached Algols,
which are particularly easy to detect (due to deep primary minima)
and are thus expected to dominate in number
over the Main Sequence stars in surveys similar to OGLE.
Of course, Algols can be recognized, even from a single
color light curve, thanks to their characteristic large
difference of depth of the minima. One can eliminate them, as we did,
by applying a criterion of equal eclipse depth, but this
introduces a restrictive limitation to
systems with components having similar effective temperatures.
On the Main Sequence, this criterion is basically equivalent
to selection of binaries with similar components, i.e.\
with mass-ratios ($q = M_2/M_1 \le 1$)
close to unity ($q \simeq 1$). Such systems may  still be
numerous; for example the shortest-period currently known
Main Sequence binary, BW3.038\footnote{The naming convention used
here is the same as in the previous papers of this series:
BW for Baade's Window (these letters are sometimes omitted),
followed by the OGLE field
number, and then the variable number, after the dot. The central
field BWC is identified by zero.}, at the very short-period
end of our distributions is such a system (Maceroni \&
Rucinski \markcite{mr97} 1997 = MR97). Nevertheless, a
limitation to the mass-ratios close to unity may be considered
a drawback in our approach. However, it was inevitable,
in view of the very limited information of single-color
light curves that we had in our disposal.

The  ``pre-contact binary'' light-curve shape filter that we used
is based on  the Fourier cosine series decomposition of the
light curves. The basic philosophy of the approach
follows the principles of the  filter used to select a sample of
contact binaries described in R97a and R97b (consult in particular
Figure~5 in R97a). The first coefficient, $a_1$,
reflects the difference between the
two eclipses and for equally-deep minima goes to zero;
the second, $a_2$ is the
largest of the coefficients and represents the total amplitude of
the light variations; $a_4$ measures the eclipse ``peakedness''
and goes to small values for the light curves of contact systems.
 As was shown in R97a, the pair ($a_2$, $a_4$) forms a powerful
separator/discriminant of  contact/detached binaries.

The  procedure  to select our sample of short period pre-contact
systems required a few  steps.
By means of the contact/detached binaries filter of R97a,
a first sample of all the
non-contact systems (339 objects) was  selected. These are the systems
falling above the ``contact line'' in the $a_4$ {\it vs.\/}
$a_2$ plane, in the left panel of Figure~\ref{fig3}.

The application of the shape filter based on the $a_1$ coefficient,
to keep only binaries with similar eclipse depths,
is shown in the right panel of the same figure.
This  step involves the selection of systems with similar components.
The equal eclipse-depth criterion implies the retention of values of
$a_1$ close to zero. A reasonable lower limit on $a_1$, and
hence a maximum allowed difference
of the minima can be decided by inspection of
the $a_1$ distributions which are shown in the two left-side
panels of Fig.~\ref{fig4}.
The lower panel shows the distribution
of the current sample of 339 detached binaries while
the upper one is for the contact binaries of the R-sample in R97a.
The R-sample is composed by W~UMa binaries
with very similar effective temperatures and thus equally-deep minima, hence
the distribution peaks at small $|a_1|$.
In contrast to the R-sample of contact binaries, the distribution of $a_1$ for
our sample of detached systems is strongly bimodal.
The comparison of the two distributions suggests that similar depths
of eclipses are selected for a limiting value of $a_1 \geq -0.017$.
This criterion establishes, however, only a necessary condition, as
$a_1$ is not only  temperature-difference dependent but also
approximately scales  with the light
curve amplitude. A filter based only on  $a_1$, would  pass, therefore,
many low-amplitude light curves for systems with low orbital inclination,
but appreciable temperature difference.
Since the second   Fourier coefficient $a_2$ is related to the light curve
amplitude, the simplest way to  take the light-curve amplitude
scaling into account is by means of the ratio $r=a_1/a_2$.
The distribution of $r$ for the same samples
of contact and detached binaries is shown in the right
panels of Figure~\ref{fig4}.   A reasoning similar to that given above for
the $a_1$ distributions suggests a threshold value of $r_{max}=0.2$.
The  selection on the depth difference has been done, therefore, using
both conditions.

A special comment is needed for the few systems with positive values
of $a_1$, and hence negative values of $r$ ($a_2$ is always negative).
By definition (see R97a),   $a_1$ is negative for a light curve where the
primary minimum is the deepest one. The OGLE team assigned
the primary eclipses to the deeper minima,
as is customary for eclipsing variables, so that $a_1$ should
in principle be always   negative.
Positive $a_1$'s can, therefore, be only due to errors introduced
either in this assignment or in the calculation of the Fourier
coefficients; the latter circumstance may result from large
photometric errors or because of a poor or uneven light-curve phase coverage.
The presence in Figure~\ref{fig4}
of several points at $a_1>0$, requires a criterion
for rejection of the positive $a_1$'s as well. Otherwise  the
poor light curves would be favored with respect to the good-quality ones.
 We decided, after a check of the light curves of the most extreme cases,
that the pure rejection of positive $a_1$ systems
would not be justified,
as some curves were indeed of rather poor quality, but of the expected shape.
The  simplest choice was therefore  to apply the previously
defined limits on $a_1$ in its absolute value.
The selection was then  performed according to:
 $|a_1|<0.017$ and $|r|<0.2$.
This additional constraint does not change
the sample in a significant way as only four systems are rejected,
but improves the consistency of the selection.

One of the four systems rejected at this stage, BW5.173,
 is a potentially interesting object  for
further studies.  It is a faint  binary ($I=17.82$) with an
orbital period $P \simeq 0.66^d$
and the light curve of a well-detached system with components of
similar effective temperatures.
After the de-reddening procedure  (see Section \ref{prop}),
it turns out, with $(V-I)_0=2.28$, to be the reddest
system of the whole sample of detached binaries, with an
absolute magnitude  $M_I=7.55$ and  a
distance $d=791$ pc. Thus, the system seems to be
very similar to BW3.038, i.e.\ it consists of a pair of M-type
dwarfs, but with a period three times longer. Since such eclipsing
systems are rare, the system is of significance as it increases the
small number of the potential calibrators of the red end of
the main sequence.

In addition to the light-curve
shape filter described above, a physical condition
on the orbital period was used as the final
step in the sample-definition to eliminate the rare
systems with evolved components. These can  be easily  recognized by
their relatively long periods, so we set an upper limit of $P=8^d$
as a reasonable value for tidally locked binaries with MS components.
The systems selected through all the criteria described above are marked
by filled symbols
in Figure~\ref{fig3}, which shows the details of our Fourier filter.

\placefigure{fig3}

\placefigure{fig4}

The sample of systems selected through the light-curve filter
and the period criterion $P < 8^d$
has been further checked for presence of systems which could
deteriorate the quality of the sample. In particular, the Fourier
coefficients for systems with partially-covered light curves may be
entirely erroneous. Visual examination of the light curves
led to removal of two systems BW2.072 and BW4.099.
After removal of six further  systems without
measured $(V-I)$ colors, the sample
consisted of 77 systems. All these, except three rejected because
of too blue intrinsic colors (see Section~\ref{prop}), formed
the final sample of 74 systems used in this paper.
The systems are listed in Table~\ref{tab1}, where -- in addition
to the original OGLE data of the period, $P$, the maximum magnitude
and color $I$ and $(V-I)$, and the amplitude $\Delta I$ -- we give the
derived values (see below): the absolute magnitude $M_I$, the de-reddened
color $(V-I)_0$ and the distance in parsecs, $d$. The Fourier
coefficients are available from the authors through their respective
Web pages\footnote{The tables of Fourier coefficients for all
933 binaries discovered by the OGLE project are located in
http://www.astro.utoronto.ca/$\sim$rucinski/ogle.html and
http://www.mporzio.astro.it/$\sim$maceroni/ogle.html}.

\placetable{tab1}

\subsection{Properties of the systems in the sample}
\label{prop}

Additional information on the properties of the systems came from
the consideration of their absolute magnitudes and de-reddened colors.
These were derived by an iterative process
of distance determination, in an approach somewhat
similar to that described in R97a, and identical to that
used in the study of the system BW3.038 (MR97).

The procedure was as follows: An adopted absolute-magnitude
calibration $M_I = M_I ((V-I)-E_{V-I}(d))$ for the main sequence was
used to find the distance $d$ and reddening $E_{V-I}$.
The procedure was iterative with the reddening allowed
to vary linearly with distance
between zero and the maximum value derived from the
background Bulge giants by Stanek \markcite{Sta96} (1996),
assuming that $E_{V-I}^{max}$ is reached at the distance of 2 kpc,
and then does not increase.
The adopted MS relation was that of Reid and Majewski
\markcite{RM93} (1993), but with a shift by 0.75
magnitude to allow for two identical stars, in consistency
with the definition of the sample of $q \simeq 1$ systems.
The results of the approach are the de-reddened colors, $(V-I)_0$,
the distances $d$ and the absolute magnitudes, $M_I$
(see Table~\ref{tab1}).

The intrinsic colors of three system, 7.004 with $(V-I)_0 = 0.21$, and
1.221 and 6.081, both with  $(V-I)_0 = 0.44$
were found to be too early for consideration in the sample of late-type
stars. The elimination of those
three systems led to the reduction of the sample to
74 systems. The intrinsic color distribution (Figure~\ref{fig5})
shows most of the
systems in the range $0.6 < (V-I)_0 < 1.3$ with a tail extending to
red colors. One may expect that M-type dwarfs populating the tail
may have different AML properties than the FGK-type
stars. Thus, we separately considered 64 systems with $(V-I)_0 < 1.3$
and 10 systems with redder colors. The border line is located
approximately at the spectral type K5V.

\placefigure{fig5}

The division into the two spectral groups was dictated not only by the
possibility of the different regime in the AML, but also because
the sampled spatial volumes are expected to be vastly different
as a function of absolute magnitudes, leading to a possibility
of very different discovery
selection effects. We can gain some insight into the matter of
the spatial depth of the samples by consideration of the increase in the
number of stars with the distance. Figure~\ref{fig6}
shows the logarithmic plots of
the cumulative number of stars versus the distance for both groups.
For the Euclidean geometry, the slope is expected to be 3, which is
approximately fulfilled by the nearby M-dwarfs. For
the FGK-type stars we see a definite deficit at small distances,
then the expected increase in number and then a strong cut-off
at about 3 kpc. The deficit at small distances is due to the bright limit
of the OGLE sample at $I=14$, while the cut-off at large distances
is due to the combination of the faint limit of the OGLE sample
and of the line-of-sight leaving the galactic disk (see
Section~9.1 in R98b). The cut-off
is relatively sharp because 72 and 92 percent of stars of the
FGK group are located closer than 3 kpc and 4 kpc, respectively.

\placefigure{fig6}

The main goal of this paper is the derivation of a statistically sound
orbital-period distribution. This distribution
may -- and probably does -- depend on the spectral type range, but with
the small number of systems we cannot avoid the necessity of grouping
the systems into relatively coarse samples. We therefore
check first if the division into only two broad spectral groups
is a legitimate one and whether the range of colors in the FGK
group is not selected too wide.

Previous studies on all-sky  samples  in the same period range
suggested a  change in the period distribution which is
dependent on the spectral type, although the division seems to be
located between early A--F and late G--M type
binaries. The distributions of about 1200 close binaries in the sky field,
grouped by spectral type of Farinella et al. \markcite{FP79} (1979)
show a change in shape from unimodal, for systems with
O--F type components, to bimodal for G~types, and to multimodal
for K--M spectral types.
A later study by Giuricin et al. \markcite{GMM84} (1984),
of 600 eclipsing and spectroscopic field binaries also shows --
though less clearly defined -- a trend towards broader distributions
for later spectral types.
These large field samples are, however, so heterogeneous
that is practically impossible to deal with the many
selection effects and  misclassifications affecting them. The current
small, but homogeneous one, provides a totally independent
and external check on these results.

\placefigure{fig7}

A first test of the homogeneity of our sample of 74 systems
was done by splitting it into color-range defined
subsamples. In this test, we tried two partitions: into
two equal-size samples of
37 systems each (containing systems respectively bluer
and redder than $(V-I)_0=0.93$),
and into two sub-samples of FGK and M  binaries, as defined above
(the division at $(V-I)_0=1.3$). The null
hypothesis of the same parent population was checked
by means of a standard two--sided Kolmogorov-Smirnov test, by
computing the maximum absolute difference between the two cumulative
distributions, $D_{KS}$, and the significance level of the
null hypothesis $P_{KS}$. The cumulative period distributions are
shown in Figure~\ref{fig7}.
We see some subtle
differences between the sub-samples, but -- taking into account the small
number of objects in the sub-samples -- they are not significant to the
point of rejection of the null hypothesis of the same period distribution.
We note that for the equal-number division, the blue sub-sample extends
from 2 to 6 kpc while the red one extends only from 0.5 to 3 kpc so that
different discovery selection biases are not excluded.
The results of the KS test, reported in Table~\ref{tab2},
do not allow firm conclusions: The significance levels for both divisions
are close to 0.6; the differences are probably mostly due to systematic
trends in discovery selection effects, but -- at least --
the results indicate that the null hypothesis cannot be rejected.

Similarly to the cumulative period distributions,
no obvious dependences are visible in the scatter diagrams
for the $(V-I)_0$ {\it vs.} $\log P$ and
$M_I$ {\it vs.} $\log P$ relations, as shown in
Figure~\ref{fig8}. This is
confirmed by the values of the correlation coefficients which are all
close to zero. The results of application of the
Kendall ($r_K$) and Spearman rank ($r_S$) correlation coefficients are given
in Table~\ref{tab2}.
We should note that although $M_I$ is partly derived from $(V-I)_0$,
it also depends on $I$ so that the correlation coefficients given
in the table do not have to be exactly same, as they appear to be.

\placefigure{fig8}

The lack of  correlation between periods and colors or absolute
magnitudes is the reason why no attempt at the definition of a volume-limited
sample (say to 3 kpc) was made, thus allowing inclusion of more
distant systems as well. We retained partition into the FGK and M groups,
however, entirely on the basis of an expectation
that the M-dwarf sub-sample may have different AML properties.
For that reason, we consider in this paper
both, the FGK group of 64 systems as well
as the full sample of 74 systems.
Obviously, the M-dwarf group is too small for any separate period-distribution
considerations.

\section{THE PERIOD DISTRIBUTION AND DISCOVERY SELECTION BIASES}
\label{dist}

\subsection{The orbital-period distribution}

We have used the sample of 64 FGK-type binaries and then the full sample
augmented by 10 M-type binaries to analyze the orbital period
distribution and thus infer the AML-driven orbital-period
evolution in the pre-contact stages.
The implicit assumption was that the
orbital periods and mass-ratios are not correlated. In the opposite case,
the pre-selection of $q \simeq 1$ systems might lead to a bias
in the period distribution.

The observed period distributions, for the whole sample
of 74 systems and for the FGK group of 64 systems, are shown in
Figure~\ref{fig9}.
Except for very short periods, $P < 0.35$ day, where no
detached FGK binaries could exist because of the onset of
contact, the data show a trend of  progressively decreasing number
of binary systems for increasing period. Weighted
fits to the histograms of $\log N$ versus $\log P$
(Figure~\ref{fig9}
in the form: $\log N = A_0 + A_1 \, \log P$, with weights
calculated on the basis of Poissonian errors in bins
$\Delta \log P = 0.2$ wide, gave
$A_0 = 1.05 \pm 0.06$, $A_1 = -0.79 \pm 0.16$
for the whole sample, and $A_0 = 1.02 \pm 0.07$, $A_1 = -0.80 \pm 0.17$
for the FGK sub-group. The values of $\chi^2$ for the fits
were 3.1 and 3.7  for the 6 log-period bins. The linear
fits are thus only marginally appropriate, but definitely
much better than flat distributions.

To define the uncertainty limits on the coefficients $A_0$ and $A_1$,
a Monte-Carlo experiment has been conducted in which weighted fits were
made to several thousand artificial Poisson distributions with the same
{\it mean\/} values for each log-period bin. Because the mean values
do not form a linear dependence, such fits provide more realistic
estimates of the uncertainties of the coefficients $A_0$ and $A_1$.
The results, expressed in terms of the distributions of
the individual determinations of the coefficients around their median
values, are given in Table~\ref{tab3}. The fits are shown in the
second panel of Figure~\ref{fig9}.

\placefigure{fig9}

The linear fits in
Figure~\ref{fig9} show  an unexpected slope
as the number of systems decreases, rather
than increases with period.
Even for a ``saturated'' AML rate (see Section~\ref{aml})
the slope should be the opposite of what is found. For
any $\alpha \ge 0$ in $\dot{H} \propto P^{-\alpha}$ the
logarithmic slope of the number distribution should be larger than 1/3,
as in $N \propto P^{\alpha + 1/3}$. In other words,
the short-period end of the period distribution for pre-contact
binaries should be always less populated.

We think that the observed trend, which is contrary to the
expectations, is entirely due
to discovery-selection bias effects scaling in proportion
to the period length. For the first time we have a well-defined
sample of eclipsing systems and for the first time we can see
the selection effects so clearly, without other discovery biases.
There are several reasons why systems with longer periods are more
difficult to detect: (1)~Chances of observing eclipses decrease with the
increasing separation of the components, as the range of
orbital inclinations  rapidly shrinks; (2)~Chances of detecting
eclipses decrease as they become progressively shorter; (3)~Fewer
orbital cycles and thus fewer eclipses
are observed for a given duration of the survey.

The only rigorous approach in handling the detection biases would be
to simulate the whole discovery process, starting
from the data-taking through all the following reduction stages.
Such simulations could be done only by the OGLE team. Since they are
not available, simplified approaches of handling the discovery
selection have been attempted.

\subsection{Orbital-period selection biases}

The discovery selection biases can be estimated
by consideration of probability that a distant observer
notices eclipses.
One way to estimate this probability is by evaluating the
solid angle subtended by the sum of the fractional radii $(r_1+r_2)/a$,
relative to hemisphere visible by a distant
observer, i.e.\ by dividing it by $2 \pi$ steradian. This relative
solid angle is given by the integral of  the eclipse
relative durations  over
the range of inclinations that can result in eclipses, which evaluates
to $1-  \left(  \sqrt{1-  \left( \frac{r_1+r_2 }{a } \right) ^2}\right)
\approx \left( \frac{r_1+r_2 }{a}\right)^2 $. For fixed radii,
the probability of discovery of an eclipsing system scales as
the inverse of its orbital separation in square, $\propto a^{-2}$.
The same proportionality is obtained by considering the
fraction of the sky one star ``sees'' covered by the other star.
Obviously, such very simple
estimates only very approximately represent trends in
the depth and duration of the eclipses, as seen by
a distant observer, but do show that the discovery selection effects
rather strongly depend on parameters of binary systems.

Below, we will take a pragmatic approach and consider other
``strengths'' (other power-law dependences)
of the discovery biases, but we feel that
the discovery-probability scaling according to $a^{-2}$
should be a particularly reasonable assumption for the equal-mass,
hence presumably equal-radius component systems that we selected.
The correction factor that we should use to multiply
the statistics of periods is then expected to behave as: $corr \propto
a^2 \propto M_{tot}^{2/3} P^{4/3}$. The mass dependence can be further
removed by observing that, as we described in Section~\ref{sample},
the orbital periods in our sample do not correlate with the color or
absolute magnitude. We can therefore assume that they do not correlate
with the total mass of the system, $M_{tot}$. The correction
factor to apply to the histograms in
Figure~\ref{fig9} would be then: $corr \propto P^{4/3}$.
Since we considered the logarithmic slopes
in fitting the $\log N$ versus $\log P$ dependences, we will from now
on -- instead of $corr$ --
use the slope correction $C$, as  in $corr \propto P^C$,
with the most likely value of $C = 4/3$.

The relation between the braking-efficiency exponent,
$\alpha$, the
observed  slope, $A_1$, and the observational selection correction,
$C$ is: $\alpha = A_1 + C  - 1/3$.
We will consider implications of
assuming various values of $C$ . For the
value of $A_1$, we have
a choice of selecting between the directly determined values of
$A_1 = -0.79$ or $-0.80$, or from the Monte Carlo experiment,
$A_1 = -0.74$ or $-0.73$. For simplicity, and because any results
will in fact be dominated
by the systematic effects characterized by $C$, we set -- from now on --
$A_1 = -0.75 \pm 0.20$ (the error estimate
comes from the Monte Carlo experiment, see Table~\ref{tab3}). Some
illustrative cases are discussed below:
\begin{description}
\item[$\mathbf C=0$:] This is a case of no discovery selection effects.
This case is hardly possible, not only because the detection biases
almost certainly
exist, but also because we obtain then a strongly inverted braking
efficiency law with $\alpha = -1.08$. This would imply that rapidly
rotating stars lose relatively less angular momentum than the slowly
rotating ones.
\item[\boldmath $\alpha = 0$: \unboldmath]
 A perfectly ``saturated'' law $dH/dt$
with no dependence on the period. The bias correction would be then
$C=+1.08$ which is only slightly less than the most preferred by
us value of 4/3. We also that evolution would in this case produce almost
no change of the initial slope (going from a value of $\beta=0.35$ to 
$\alpha+1/3=1/3$).
\item[$\mathbf C=+4/3$:] This is our preferred value of the bias
resulting in the value of $\alpha = +0.25$, i.e.\ close to
saturation, yet with a weak
acceleration of the AML evolution with shortening of the period.
\item[$\mathbf C > +1.75$:] This inequality is considered
here following the theoretical arguments of St\c epie\'n \markcite{ste95}
(1995) that $\alpha - 2/3 > 0$ (see his Figure~1), (a particular  value 
in this range is 
the local fit of the braking law, as used in Figure~\ref{1} that yields 
$C=2.57$). These  solutions
would imply that the current sample suffers from  stronger
discovery selection biases than expected for $C = 4/3$.
\end{description}

In conclusion, we feel that the most likely value of the
AML braking-law exponent is close to
$\alpha \simeq 0$. The discovery selection effects
are strong and important, yet the departure from the bias exponent
$C \simeq 4/3$
would imply implausible combinations of parameters.
For $C=0$, the period distribution would have a genuinely
negative slope, with more binaries accumulating at
the short-period end of the period distribution, just before the
conversion into contact binaries.
We note that such a law would agree with the
statistics of short-period systems in Hyades (Griffin \markcite{griff85}
1985; see Figure~\ref{fig2} in Section~\ref{aml}),
provided, of course, that it does not have its own detection biases.
However, it is really hard to imagine that
the discovery biases do not exist in the OGLE database, so that
almost certainly $C \ge 0$. Applying progressively larger
corrections $C$ would make $\alpha$ closer to zero and then positive.
If $C>4/3$, then the discovery biases would be exceptionally strong.
We cannot exclude this possibility, but consider it unlikely.

The referee of the first version of the paper pointed out
that a saturated braking law extending to periods as
long as our upper limit may be in disagreement with the results
for late-type single stars.  The activity indicators that can be related
to the AML  rate show a change of slope -- from a saturated to a steeper
law
-- at  rotation periods around  3--4 days (see for instance Wichmann
et al.
\markcite{Wic98} 1998). A mass-dependent slope transition
(from $P \approx  2^{d}$ for $m=m_\odot$  to $P \approx  9^{d}$
for $m=0.5 m_\odot$) was included in the theoretical
models of Bouvier et al.\ \markcite{Bou97} (1997) and found
to provide a good fit of PMS and MS rotational data.
The small size of our sample and the consequent relatively coarse
binning
did not allow us to properly analyze this feature. The expected
change of the slope would fall in the tail of our distribution since
we currently have only three systems in the bin $P>5$ days.
On a qualitative ground, Figure~\ref{fig9} does suggest that an
exclusion of the last bin from the fit would result in a slightly
flatter
braking law, but we feel that an estimate of the slope based
on such reduced data would actually be an over-interpretation of the
available material.
 
\section{DISCUSSION AND CONCLUSIONS}
\label{disc}

We have used the by-product of the OGLE microlensing
project, the database
of eclipsing binaries toward Baade's Window, to analyze the period
distribution of short-period ($0.19 < P < 8$ days), late-type,
main-sequence systems.
This distribution was used to shed light on the angular-momentum-loss
efficiency for rapidly-rotating, late-type stars. The final results
are very tentative as they are
totally dominated by the systematic effects of discovery biases.
Yet, they may be of importance for similar future applications of
microlensing databases, so that certain lessons can be learned from
our experiences.

The main, obvious lesson is thus a very high importance of testing the
variability-discovery algorithms for period-length biases. This
can be done only by the observing teams,
by simulations of detectability
of variable stars. Any a posteriori corrections to the
observed statistics have debatable value, as is the case with our
slope correction $C$,
introduced in Section~\ref{dist}. Depending on its value,
we can obtain any braking law $dH/dt \propto P^{-\alpha}$, with
$\alpha = -1.1 + C$. We tend to favor a relatively strong
detection bias law described by $C = +4/3$ and  (implying a
braking law close to the ``saturated'' one of $\alpha \simeq 0$),
but any value of $C$ is basically permissible. It seems very unlikely
that $C = 0$, that is that the OGLE database has no period biases.
For this case, however, the negative slope of the braking law
would agree with the statistics of short-period binaries in Hyades
(Griffin \markcite{grif85} 1985), with more binaries at the short-period
end.

The other lesson is the availability of color curves.
Because the light curves were observed in one color only, we had
to use a very strong light-curve-shape criterion to filter out
the easily-detectable Algols and define a sample of short-period,
main-sequence systems with almost identical components.
The sample that we defined consists of
74 objects, but would be larger if all systems with MS components
having un-equal temperatures were not rejected.
Because of this restriction, we could not meaningfully consider any
quantities related to spatial frequency of occurrence
(such as luminosity
or period functions discussed extensively for the OGLE contact systems
in R98b). This resulted in a limitation that the
available quantity was the slope of the period distribution,
and even that
affected by strong discovery selection biases.

We note that
our sample of 74 systems consisted of 64 binaries of spectral types
F, G and K and 10 M-type systems with a large spread in colors.
We expected differences in properties between these samples and
thus considered them separately, but did not notice any major
disparities which would affect the period-distribution statistics.

In addition to these somewhat general statements,
we add a minor, but firmer conclusion that
there exists no evidence in our homogeneous sample for any bimodal
period distribution, similar to that
found by Farinella et al.\ \markcite{FP79} (1979)
with a dip at $\log P \simeq 0.2$.
We feel that this dip and a secondary maximum in the distribution
were most probably produced by misclassification
of evolved systems as consisting of main-sequence components. A
smooth period distribution, without any structure,
implies a different form of the braking law with respect to
that derived by MV91 and Maceroni \markcite{cm92} (1992).
Though the functional form remains similar, in the sense of
a ``saturated'' braking at short orbital periods ($\alpha=0$),
the other main feature, namely  the sudden increase of the
braking efficiency for rotation of $\sim 10 \omega_{\odot}$
is no longer mandated by the data. That feature
was needed to reproduce
the bimodal distribution of field G-type systems,
that had a pronounced dip around $\log P =0.2$ days in the
distribution of Farinella et al.\ \markcite{FP79} (1979).

\acknowledgements
We would like to acknowledge the OGLE team for
the access to their database.

This work was partially supported by
research grants to CM of the Italian MURST (Ministery of University,
Scientific and Technological Research)
and of the Italian Space Agency. This work was started by
SMR during his employment by the Canada-France-Hawaii Telescope.

\clearpage 

\noindent
Captions to figures:

\figcaption[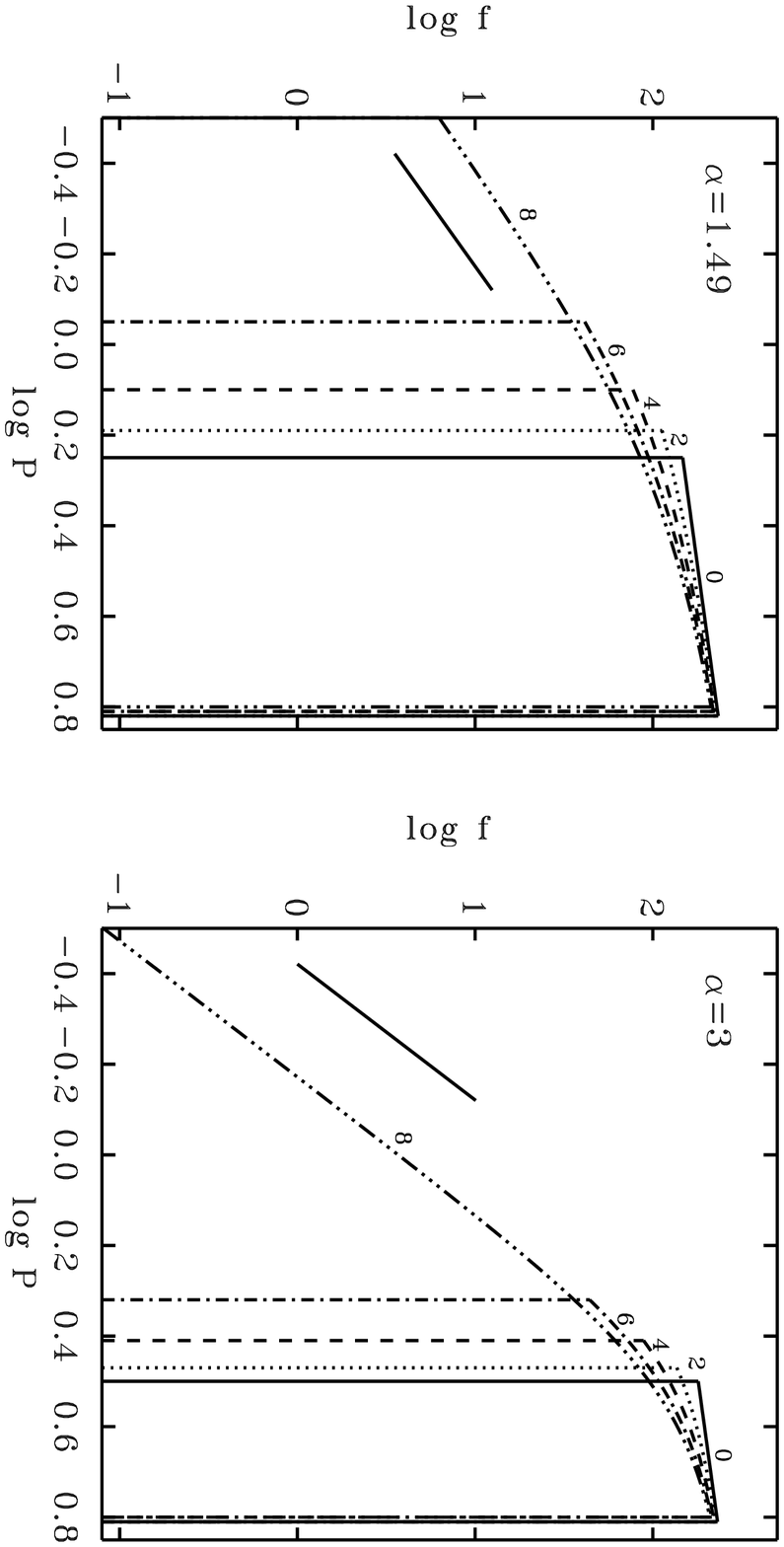] {\label{fig1}
 The figure shows the expected evolution of the period distribution for
two values of $\alpha$, 1.49 and 3, following the theoretical
predictions of St\c epie\'n  (1995) and 
Skumanich  (1972). 
The starting
distribution in both cases have the same, gently-positive slope $\beta =
0.35$, as in $f_0 \propto P^\beta$. The initial period ranges are
different for each panel; they have been adjusted to populate the
distribution down to 0.3 days after the time interval of 8 Gyrs. The
numbers by the lines give the ages of the repective populations in Gyr.
The short vectors in the left parts of the panels give the asymptotic
slopes $\alpha + 1/3$, as expected from considerations in the text.
}

\figcaption[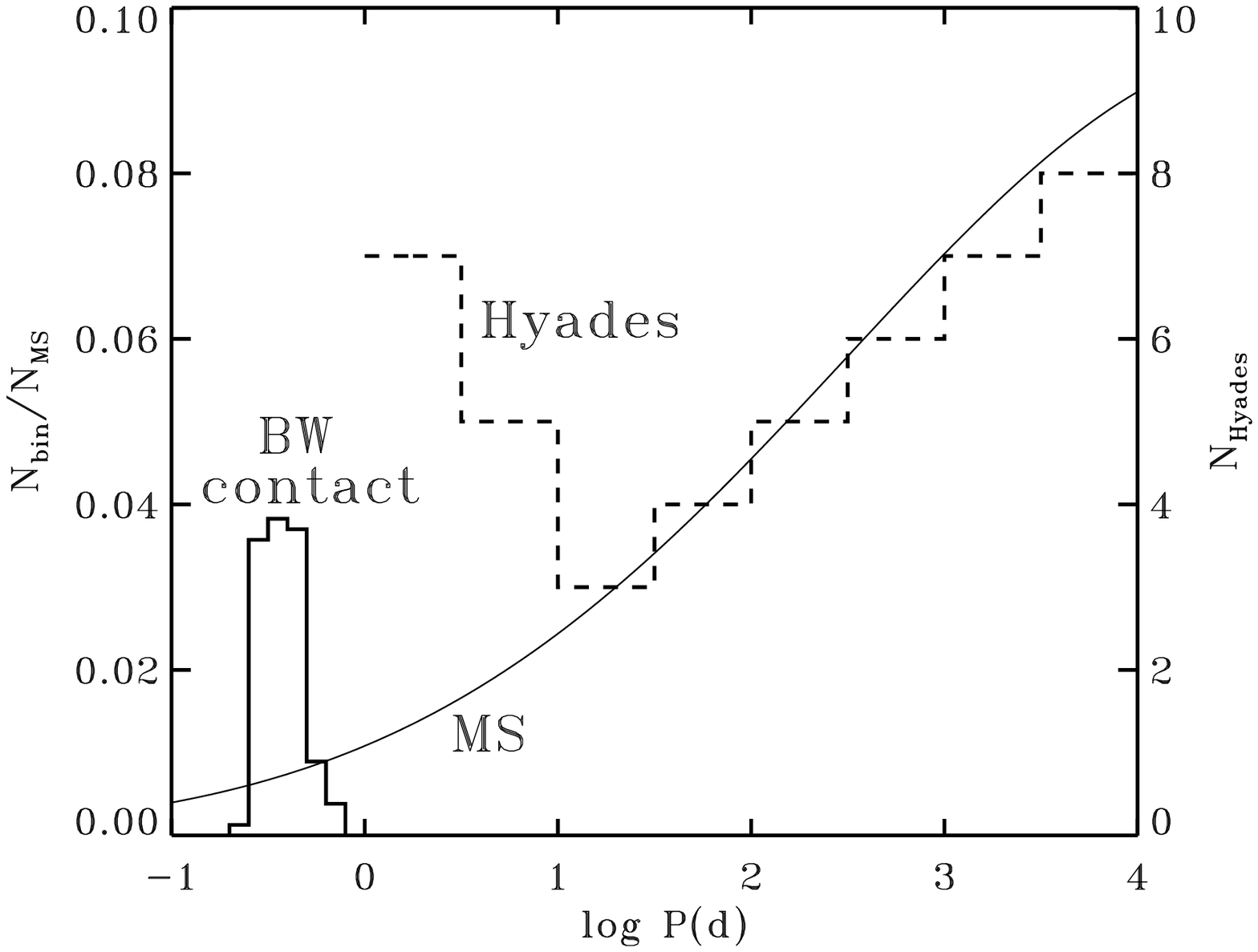] {\label{fig2}
The observed period distribution for contact binaries of the disk
population (the solid-line histogram and left vertical axis),
based on the R98b results and
the data for Hyades dwarfs (Griffin 1985; the broken-line histogram
and right vertical axis) are compared here
with the shape of the extrapolated DM91 distribution of
DM91 (thin-line curve).
Note that the statistics for the contact binaries
is based on about one hundred objects (the long-period side of the
peak is actually based on more than 200 objects), whereas the
number of short-period binaries observed in Hyades is a dozen or so,
so that the statistics is quite uncertain.
}

\figcaption[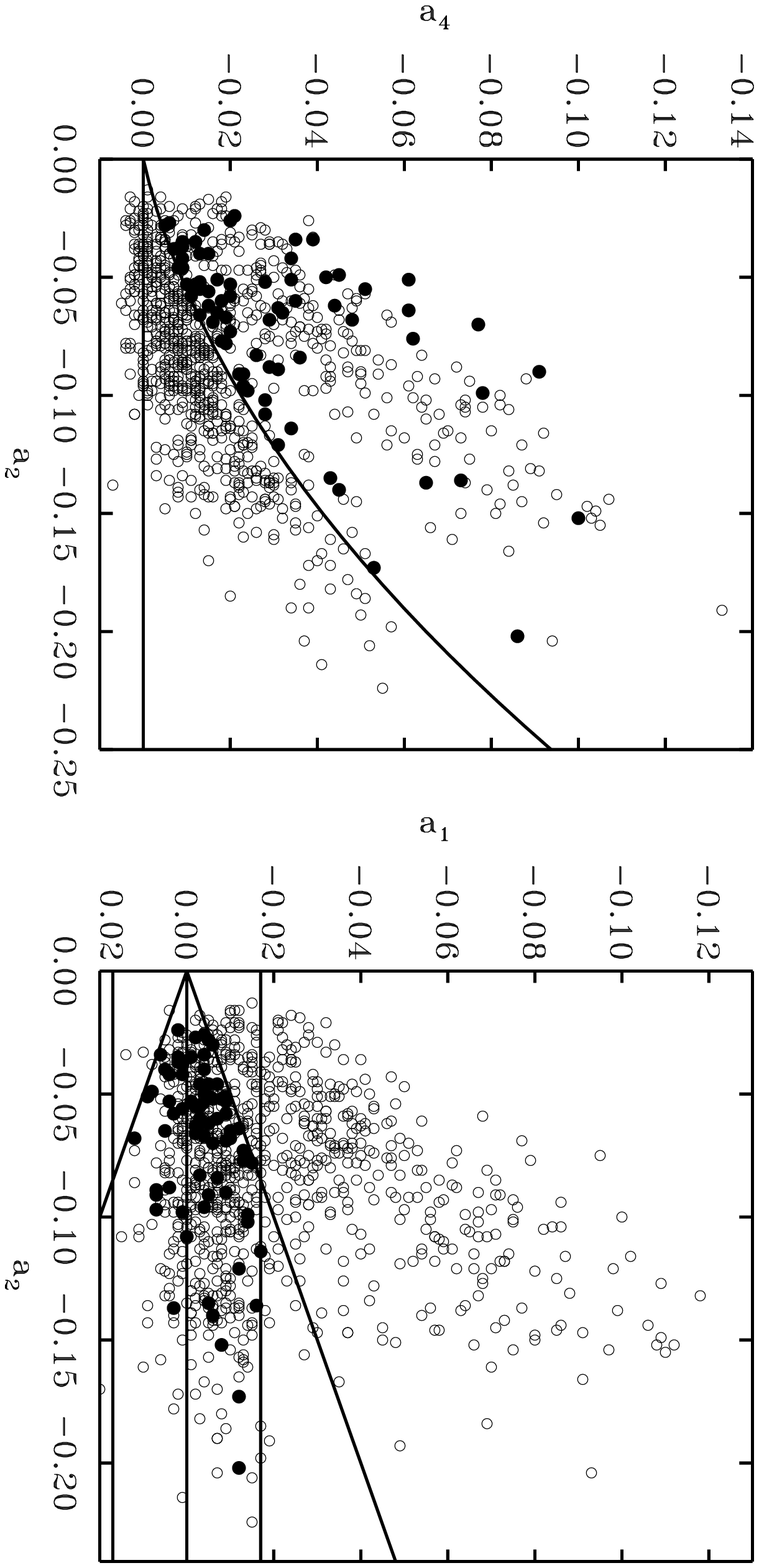] {\label{fig3}
The Fourier-coefficients $a_1$, $a_2$ and $a_4$ formed the
main part of the shape filter utilized to select short-period
pre-contact systems, as described in the text. The systems
accepted for the final sample are marked by filled symbols.
}

\figcaption[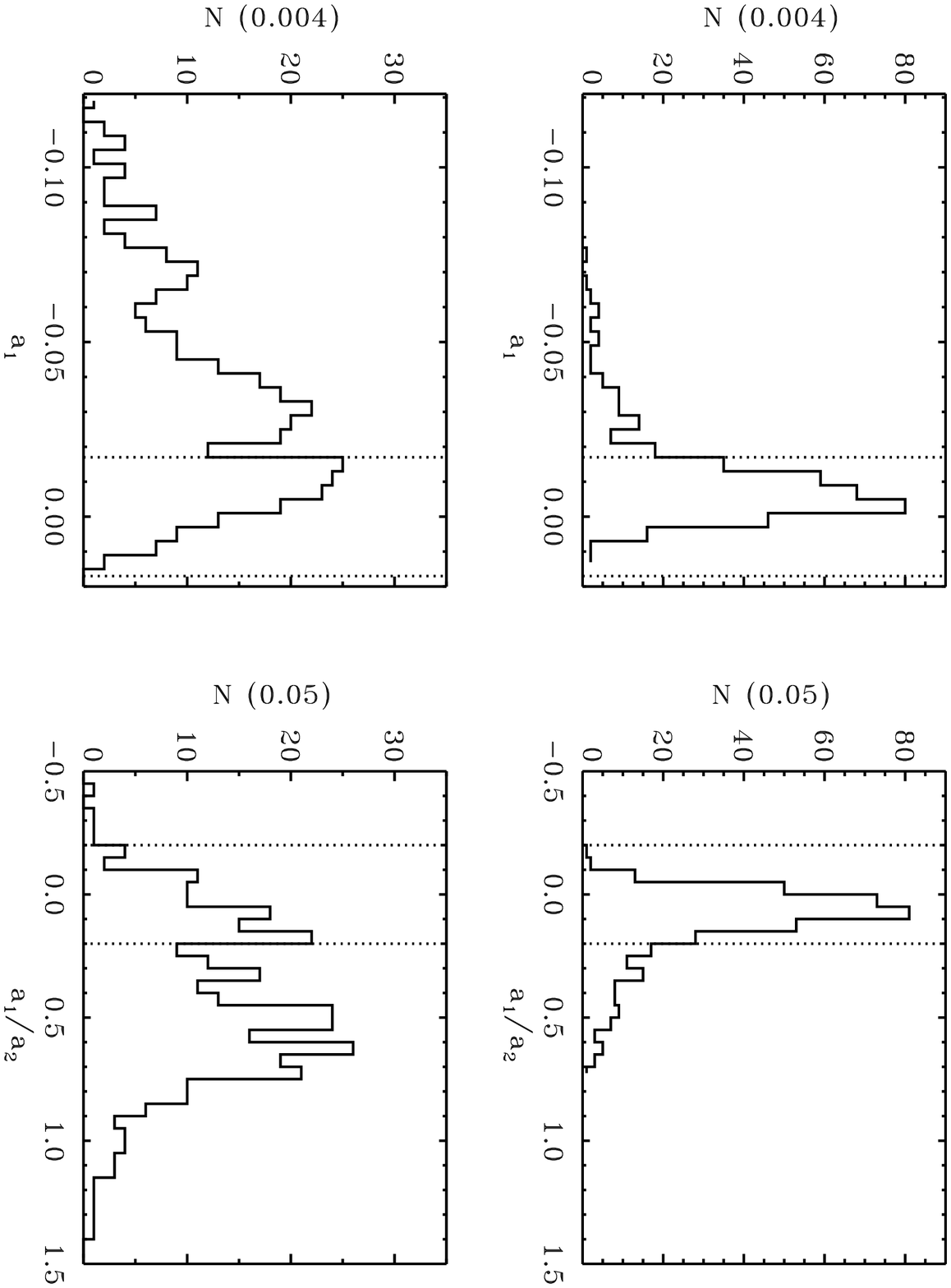] {\label{fig4}
The distributions of $a_1$ (left panels) and $a_1/a_2$
(right panels) for two samples: for 339 detached binaries
(lower panels), and for the R-sample of W~UMa contact binaries as
selected in R97a (upper panels) are shown here to illustrate
different properties of the contact and detached binary samples.
The dotted lines show the limits that have been chosen
by comparison between the distributions of the two
samples  to select  binaries with equal-temperature (and,
presumably, equal-mass) components.
}

\figcaption[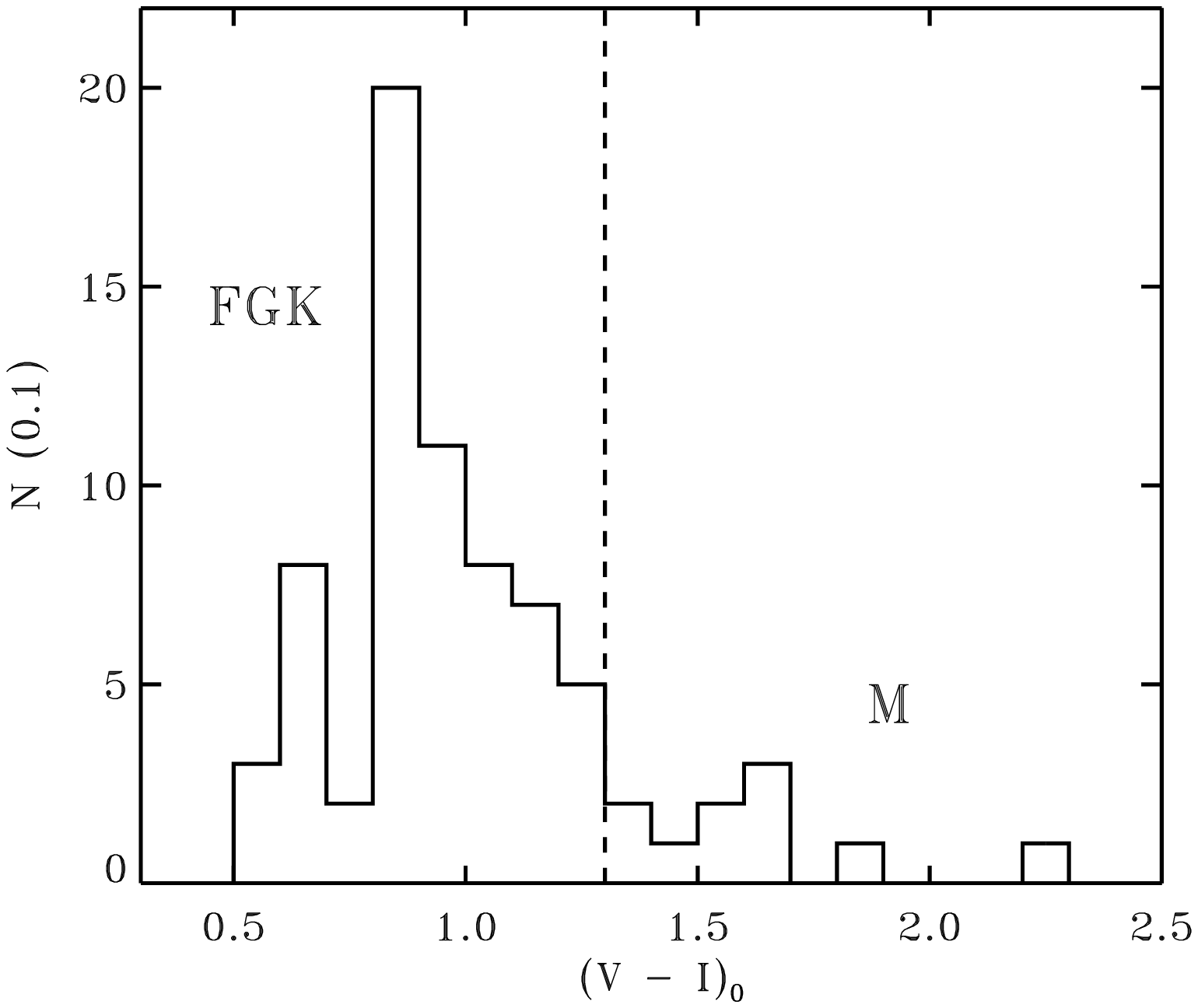] {\label{fig5}
The color distribution for the whole sample of pre-contact
systems. The border between the two spectral groups was fixed at
$(V-I)_0 = 1.3$.
}

\figcaption[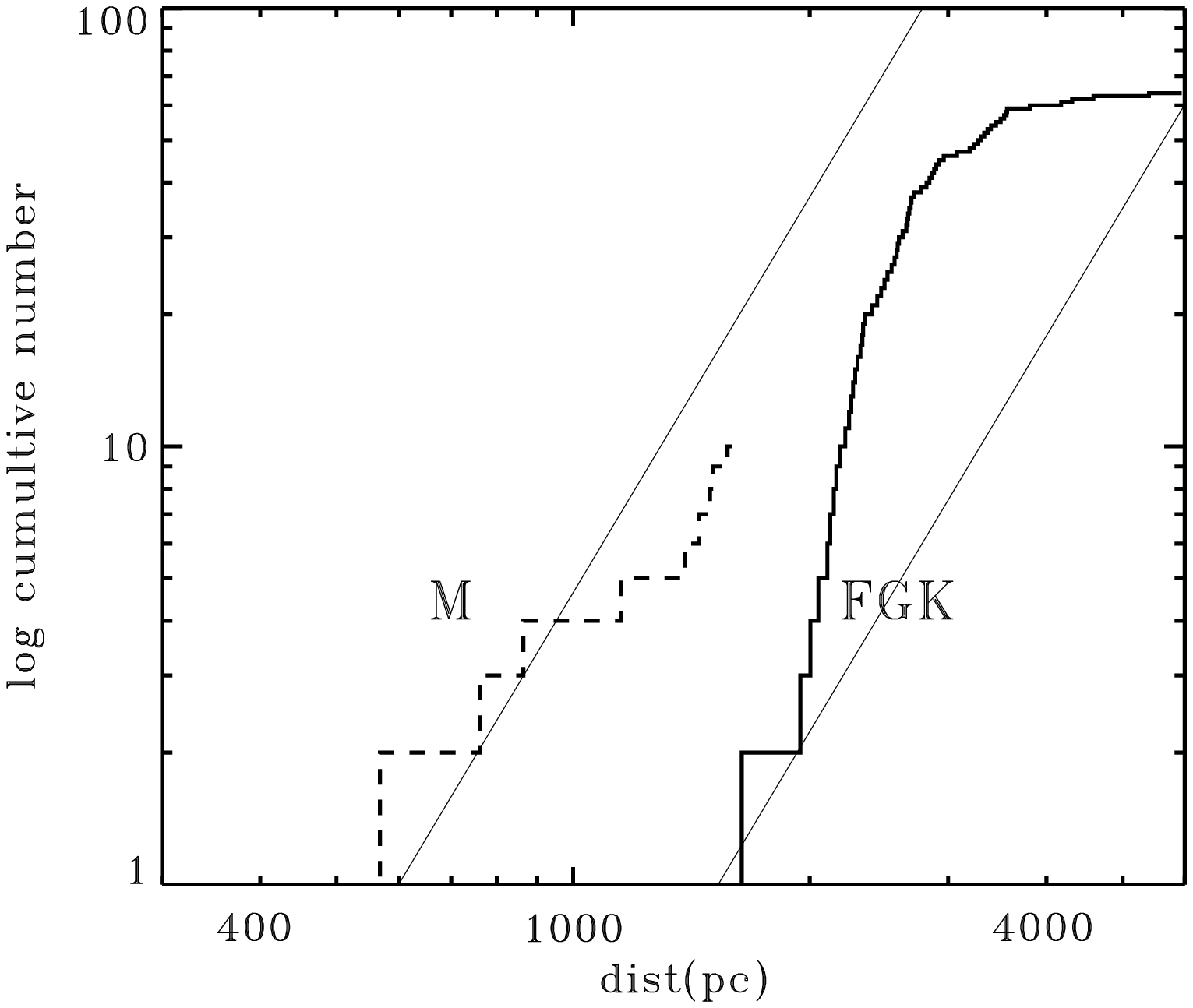] {\label{fig6}
The cumulative number distributions with distance of
systems for the FGK and M groups are shown by
full-line and broken-line histograms. The thin lines
give the uniform density relations with the Euclidean slope of 3.
}

\figcaption[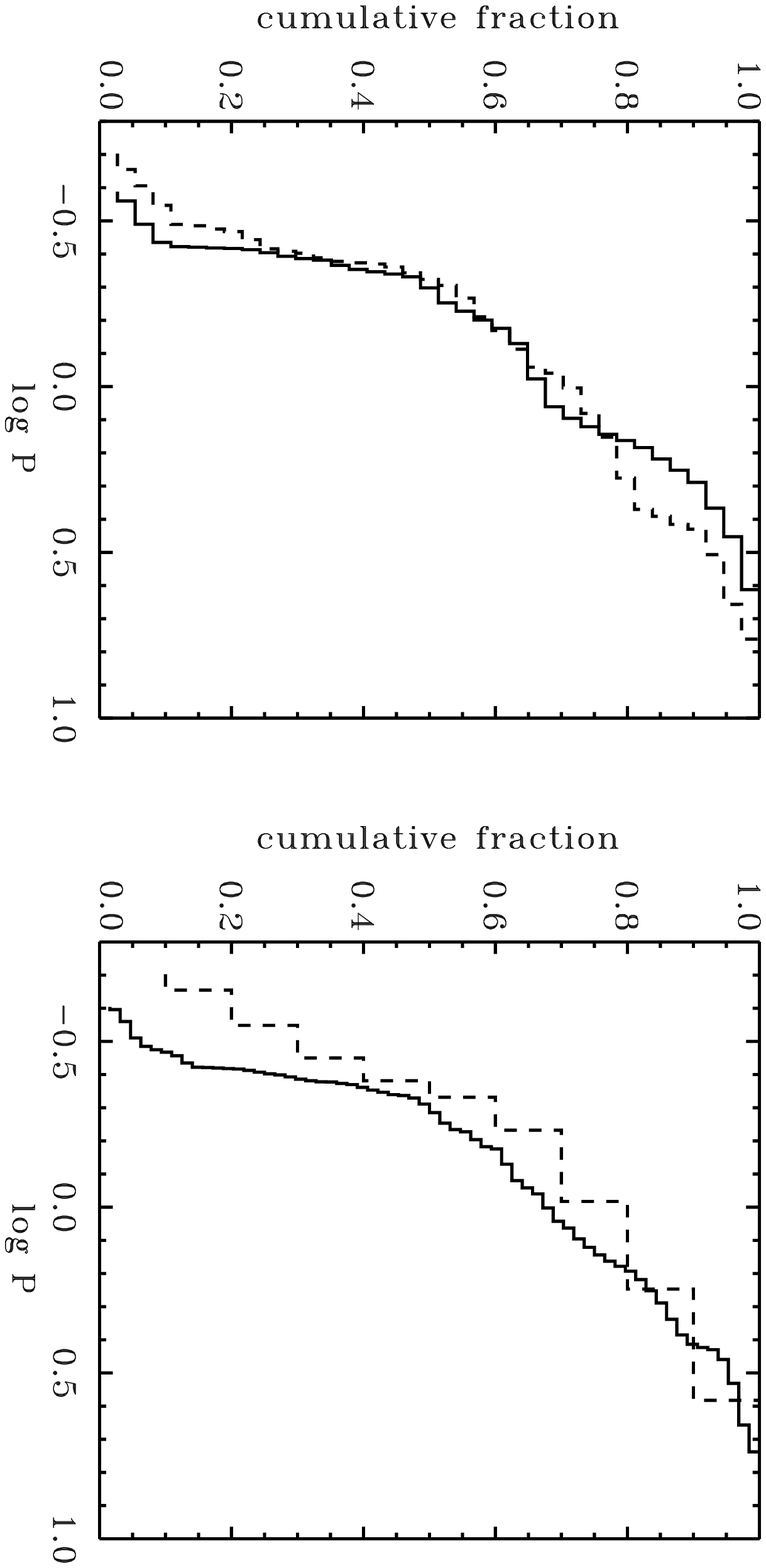] {\label{fig7}
Left panel: the cumulative  relative distribution with period of
two equal size samples of 37 systems grouped by color,  systems
bluer (full line) and redder (broken line) than $V-I=0.93$.
Right panel:  the  cumulative relative distribution of
64 FGK and  10 M-type systems (full and broken line). 
}

\figcaption[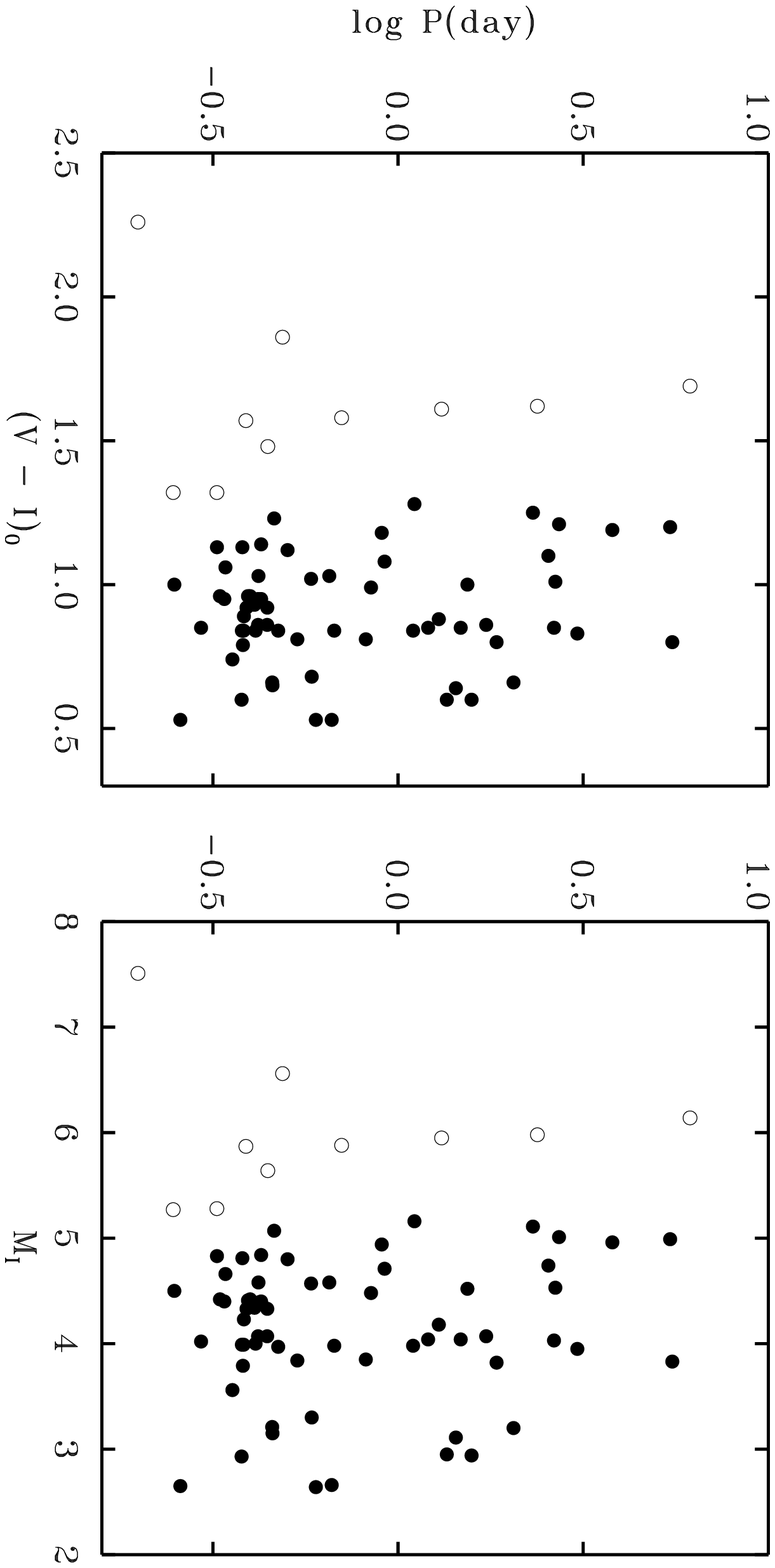] {\label{fig8}
The scatter diagrams illustrating lack of correlation between the
orbital period and the color (left panel) or the absolute magnitude
(right panel). The two spectral groups are marked by filled (FGK sample)
and open (M sample) symbols. Note that the absolute magnitudes are
for the combined brightness of the components  which were
assumed to be identical.
}

\figcaption[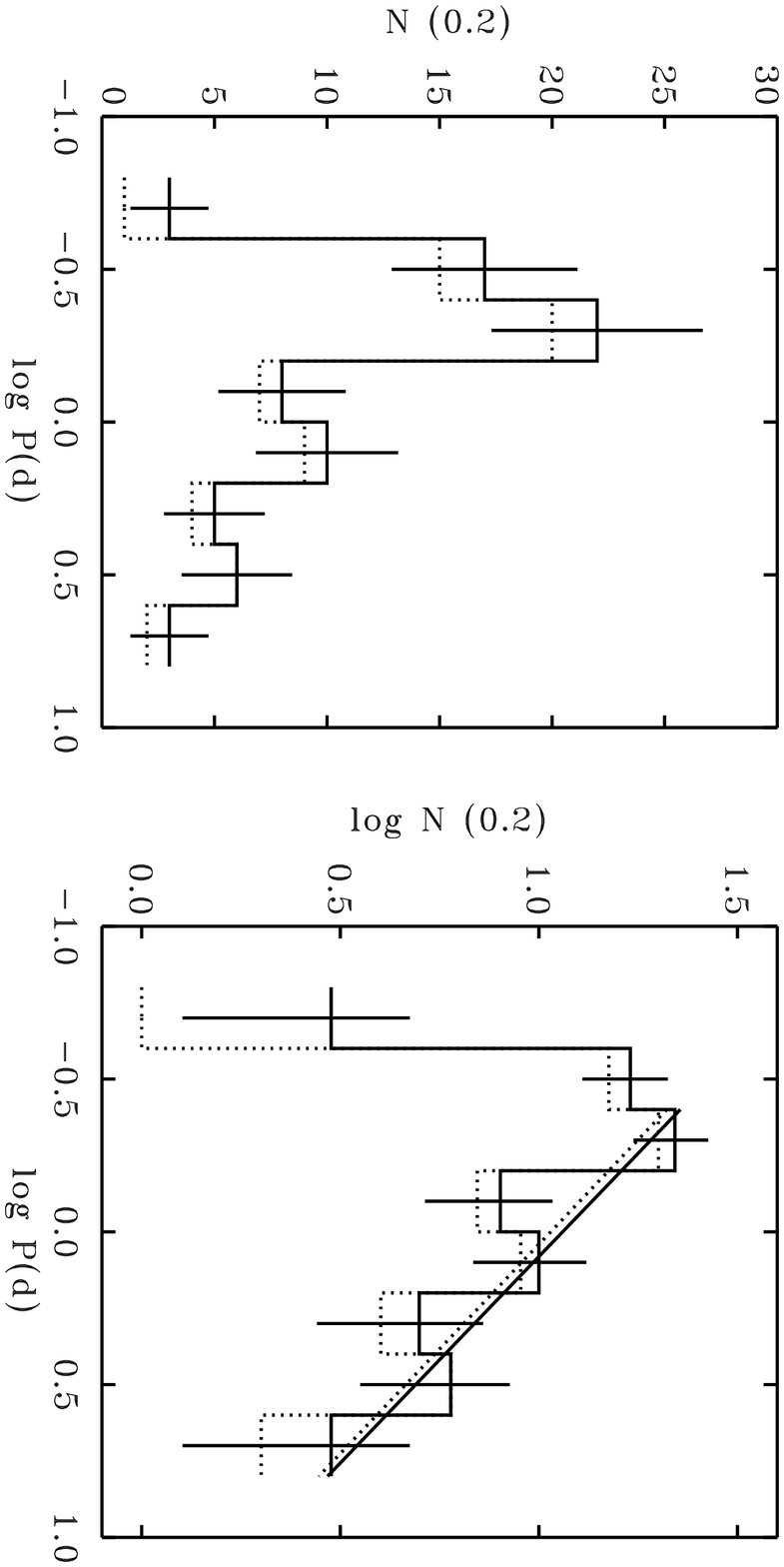] {\label{fig9}
The orbital period statistics in linear (left panel) and
logarithmic (right panel) units of the number of systems in
bins $\Delta \log P = 0.2$. The histograms are shown for
the full sample of 74 systems (solid lines)
and the FGK group of 64 systems (dotted line), but -- for clarity --
the Poisson errors are shown as vertical bars for the
full sample data only. The sloping lines in the logarithmic (right
hand side) panel show the linear fits, as described in the text.
}

\clearpage

\begin{table}
\dummytable  \label{tab1}
\end{table}
\begin{table}
\dummytable \label{tab2}
\end{table}
\begin{table}
\dummytable \label{tab3}
\end{table}

\end{document}